# Zero-Shot Surgical Tool Segmentation in Monocular Video Using Segment Anything Model 2

[1,*]Ange Lou, [2,*]Yamin Li, [2,*]Yike Zhang, [3]Robert F. Labadie, [1,2]Jack Noble

[1]Department of Electrical Engineering, Vanderbilt University
[2]Department of Computer Science, Vanderbilt University
[3]Department of Otolaryngology – Head & Neck Surgery, Medical University South Carolina
*Co-first authors
{ange.lou, yamin.li, yike.zhang, jack.noble}@vanderbilt.edu labadie@musc.edu

**Abstract**: The Segment Anything Model 2 (SAM 2) is the latest generation foundation model for image and video segmentation. Trained on the expansive Segment Anything Video (SA-V) dataset, which comprises 35.5 million masks across 50.9K videos, SAM 2 advances its predecessor's capabilities by supporting zero-shot segmentation through various prompts (e.g., points, boxes, and masks). Its robust zero-shot performance and efficient memory usage make SAM 2 particularly appealing for surgical tool segmentation in videos, especially given the scarcity of labeled data and the diversity of surgical procedures. In this study, we evaluate the zero-shot video segmentation performance of the SAM 2 model across different types of surgeries, including endoscopy and microscopy. We also assess its performance on videos featuring single and multiple tools of varying lengths to demonstrate SAM 2's applicability and effectiveness in the surgical domain. We found that: 1) SAM 2 demonstrates a strong capability for segmenting various surgical videos; 2) When new tools enter the scene, additional prompts are necessary to maintain segmentation accuracy; and 3) Specific challenges inherent to surgical videos can impact the robustness of SAM 2.

## 1. Introduction

The rapid development of the computer vision field has seen foundation models demonstrating impressive zero-shot and few-shot capabilities across various tasks. Notable examples include the Segment Anything Model (SAM) [1] for semantic segmentation, Depth Anything [2] for pixel-wise depth map prediction, and Mesh Anything [3] for mesh generation. Among these models, Vision Transformers (ViT) [4] have shown exceptional ability in learning general representations from large datasets.

Tracking surgical tools in videos is a crucial task for understanding surgical scenes and reconstructing dynamic surgical environments. Accurate segmentation of different tools is essential, but obtaining pixel-level labels for large amounts of data is resource-intensive. While semi-supervised methods [5] can significantly reduce labeling time, they still require hundreds of annotations, and the complexity of the scene can further increase this burden.

The Segment Anything Model (SAM) was the first foundation model released for semantic segmentation and has demonstrated promising results across various domains. However, when segmenting video data, it still requires prompts for each frame, which can be time-consuming and impractical for dynamic scenes.

Recently, the Segment Anything Model 2 (SAM 2) [6] has extended the zero-shot segmentation capabilities of the original SAM to video data. Trained on the SA-V dataset, which includes 35.5 million masks across 50.9 thousand videos, SAM 2 demonstrates robust zero-shot abilities for video segmentation. Additionally, SAM 2 incorporates a memory bank that facilitates the propagation of prompts from the first frame throughout the video. This feature makes it particularly well-suited for the segmentation and tracking of surgical tools in surgical videos.

In this study, we first assess the zero-shot segmentation performance of the SAM 2 model on different surgery type, including endoscopy and microscopy, as well as different surgical scenarios, including multiple tools and various video lengths.

## 2. Experiments and Performance

**Endoscopy surgery dataset.** For evaluating the performance of SAM 2 in endoscopic surgery, we selected three public datasets: EndoNeRF [7] , EndoVis'17 [8], and SurgToolLoc [9]. The EndoNeRF dataset includes two surgical video clips containing 63 and 156 frames, respectively. The EndoVis'17 dataset comprises 8 robotic surgical videos, each with 255 frames and corresponding ground truth segmentation masks. Additionally, the SurgToolLoc dataset consists of 24,695 video clips, each lasting 30 seconds and captured at 60 frames per second (fps). All these endoscopic surgery datasets were obtained from the da Vinci robotic surgical system.

**Microscopy surgery dataset.** To qualitatively evaluate the performance of SAM 2 in microscopy surgery, we selected two surgical cases from our cochlear implant dataset from Vanderbilt University Medical Center and Medical University South Carolina. These cases vary in length, ranging from 2 to 10 seconds, and encompass different surgical phases, including drilling and implant placement.

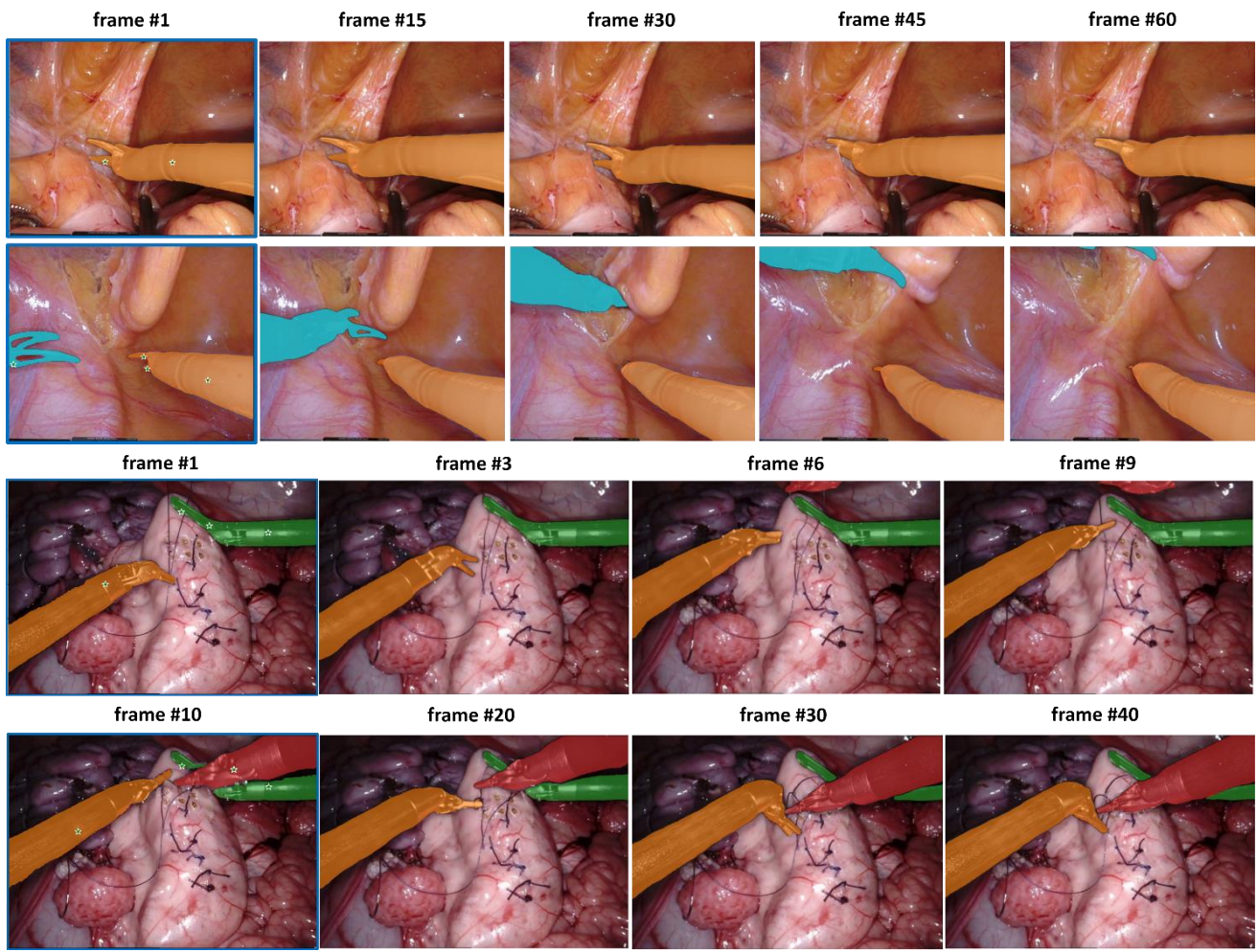

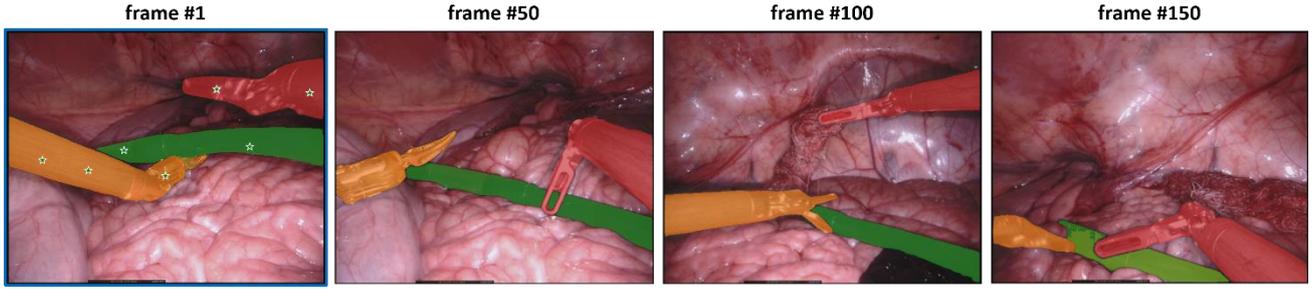

Figure 1. Results from the endoscopy surgery datasets. From top to bottom, the results are shown for the EndoNeRF, SurgToolLoc, and EndoVis'17 datasets. The first column of images represents the frames where manual prompts were applied.

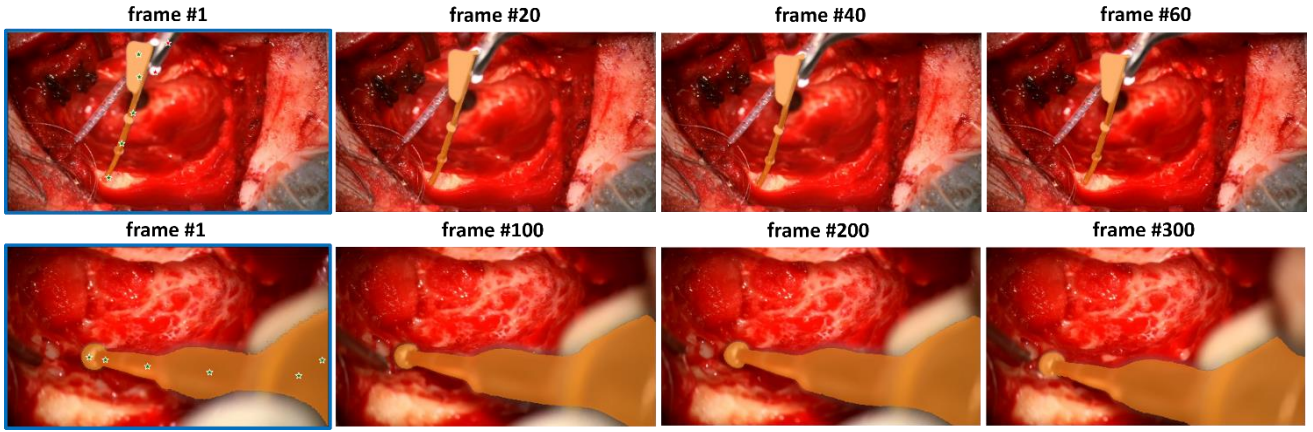

Figure 2. Results from the microscopy dataset. Two cases from our cochlear implant dataset are shown. The top row represents the implant placement phase (suction tube in orange and cochlear implant electrode in aqua), and the bottom row represents the drilling phase (drill in orange).

Table 1. Quantitative results on EndoVis' 17 dataset.

| Method | Dice ↑ | IoU ↑ | MAE ↓ |
|---|---|---|---|
| U-Net [10] | 0.894 | 0.840 | 0.027 |
| UNet ++ [11] | 0.909 | 0.841 | 0.026 |
| TransUNet [12] | 0.904 | 0.826 | 0.029 |
| **SAM 2** | **0.937** | **0.890** | **0.018** |

Table 2. Results of Paired t-Test Comparing SAM 2 and Fully Supervised Methods.

| Pairs | Dice | | IoU | | MAE | |
|---|---|---|---|---|---|---|
| | t-value | p-value | t-value | p-value | t-value | p-value |
| U-Net/SAM 2 | -7.489 | <0.001 | -8.651 | <0.001 | 5.190 | <0.001 |
| UNet ++/SAM 2 | -3.690 | <0.001 | -4.329 | <0.001 | 3.200 | 0.002 |
| TransUNet/SAM 2 | -5.494 | <0.001 | -6.489 | <0.001 | 4.008 | <0.001 |

## 3. Results

Qualitative segmentation performance of SAM 2 can be observed in Figure 1 on the 3 endoscopy and Figure 2 on the microscopy datasets. As seen in the figures, when the surgical scene has good illumination conditions and high-quality motion of the surgical tools, SAM2 can provide robust segmentation performance for both single and multiple objects.

Since the EndoVis dataset has ground truth segmentations available for the da Vinci tools, we were able to assess quantitative performance of SAM 2 compared with other state-of-the-art segmentation methods (Table 1). As seen in the table, SAM 2 outperforms U-Net, UNet++, and TransUnet in segmentation accuracy in terms of Dice score, IoU, and MAE. Statistically significance was assessed using paired t-tests. We also reported the results of paired t-tests in Table 2, which show significant improvement compared to the fully supervised segmentation method.

## 4. Discussion and Conclusion

The overall performance of the SAM 2 model is promising, even when only point prompts are provided in the first frame of the surgical video. However, several limitations need to be addressed in future work. Firstly, the model's performance tends to degrade when processing long video sequences. As illustrated in the drilling phases of Figure 2, SAM 2 loses the fine details of the segmentation of the drilling tools around frame #300. This issue presents a significant challenge for real-time, accurate surgical tool segmentation applications, where streaming videos are common.

Moreover, the surgical environment significantly impacts the model's overall efficiency. Factors such as scene blurriness, patient bleeding, and frequent occlusions can adversely affect the accuracy of surgical tool segmentation. Blurriness often results from camera motion or out-of-focus shots, while bleeding and occlusions obscure the visual cues necessary for tracking surgical tools. In our cochlear implant cases, as shown in Figure 2, challenges from the microscope camera compromise video quality, and the interaction between the tool and the surgical surface causes SAM 2 to lose precision. These factors contribute to suboptimal overall segmentation performance.

The issues mentioned above can be partially addressed by using additional prompts to enhance model performance and ensure reliable surgical tool segmentation in diverse and complex surgical scenarios. For instance, in cases from the SurgToolLoc dataset (Figure 1), introducing new tools into the surgical environment can benefit from additional prompts to maintain segmentation accuracy.

Notably, the SAM2 model achieves good performance even under zero-shot evaluation, demonstrating its promising generalization capabilities. SAM2 offers a series of pre-trained weights suitable for different model scales. Qualitative and quantitative comparisons of surgical videos with varying model sizes will be included in the full paper and all video results will be available in external link: https://github.com/AngeLouCN/SAM-2_Surgical_Video.

Future work should focus on improving the quality of long video sequences and fine-tuning the SAM2 model for specific tasks to mitigate the adverse effects of challenging environmental conditions. Addressing these limitations is crucial for the practical deployment of SAM2 in clinical settings, ensuring its reliability and effectiveness in assisting surgeons during operations.

## New and breakthrough work to be presented

In this study, we present the first evaluation of the Segmentation Anything Model 2 (SAM 2) on surgical videos. Using the extensive Segmentation Anything Video (SA-V) dataset, SAM 2 demonstrates impressive zero-shot performance, effectively segmenting surgical tools with minimal prompts. Our work highlights the model's robustness in various surgical scenarios, providing valuable insights for future improvements of SAM 2 in the surgical domain and enhancing its potential for real-time applications in clinical environments.

## Acknowledgements

This work was supported in part by NIH grant R01DC008408 from National Institute of Deafness and Other Communication Disorders. The content is solely the responsibility of the authors and does not necessarily reflect the views of this institute.